\documentclass[aps,prd,twocolumn,groupedaddress,preprintnumbers,superscriptaddress]{revtex4-2}  
\usepackage{graphicx,mathtools,tabu,array}
\usepackage{epstopdf}
\usepackage{amsmath}
\usepackage{amsfonts}
\usepackage[colorlinks=true,citecolor=red,linkcolor=blue,breaklinks=true]{hyperref}
\usepackage{accents}
\usepackage[figure, table]{hypcap}
\usepackage{amssymb}
\usepackage{appendix}
\usepackage{comment}
\usepackage{bbold}
\usepackage{xcolor}
\usepackage{slashed}
\usepackage{subfigure}
\usepackage{setspace}
\usepackage{footnote}
\usepackage{multirow}
\usepackage{braket}
\usepackage[normalem]{ulem}
\usepackage[utf8]{inputenc}
\usepackage{mathrsfs}
\usepackage{longtable}
\usepackage{enumitem}
\usepackage{orcidlink}
\usepackage{booktabs} % Enhance the tables
\usepackage{subfigure}

\newcommand{\PreserveBackslash}[1]{\let\temp=\\#1\let\\=\temp}
\newcolumntype{C}[1]{>{\PreserveBackslash\centering}p{#1}}
\newcolumntype{R}[1]{>{\PreserveBackslash\raggedleft}p{#1}}
\newcolumntype{L}[1]{>{\PreserveBackslash\raggedright}p{#1}}

\LTcapwidth=\textwidth

\DeclareMathAlphabet{\mathpzc}{OT1}{pzc}{m}{it}

\graphicspath{{Figures/}}

\allowdisplaybreaks

\definecolor{palatd}{RGB}{104, 36, 109}
\definecolor{palatb}{RGB}{0, 56, 168}
\definecolor{palatr}{rgb}{0.745,0.118,0.176}
\newcommand\myshade{80}
\colorlet{mylinkcolor}{palatr}
\colorlet{mycitecolor}{palatb}
\colorlet{myurlcolor}{palatd}

\hypersetup{
  linkcolor  = mylinkcolor!\myshade!black,
  citecolor  = mycitecolor!\myshade!black,
  urlcolor   = myurlcolor!\myshade!black,
  colorlinks = true
}

\begin{document}
\sloppy  

\preprint{IFT-UAM/CSIC-26-02, TIFR/TH/26-5}

\title{Emergent Large Lepton Mixing from Neutrino Refraction in Dark Matter}

\author{Susobhan Chattopadhyay\,\orcidlink{0009-0000-2346-2273
}}
\email{susobhan.chattopadhyay@tifr.res.in}
\affiliation{Tata Institute of Fundamental Research, Homi Bhabha Road, Colaba, Mumbai 400005, India}

\author{Yuber F. Perez-Gonzalez\,\orcidlink{0000-0002-2020-7223}}
\email{yuber.perez@uam.es}
\affiliation{Departamento de F\'{i}sica Te\'{o}rica and Instituto de F\'{i}sica Te\'{o}rica (IFT) UAM/CSIC, Universidad Aut\'{o}noma de Madrid, Cantoblanco, 28049 Madrid, Spain}

\author{Manibrata Sen\,\orcidlink{0000-0001-7948-4332}}
\email{manibrata@iitb.ac.in}
\affiliation{Department of Physics, Indian Institute of Technology Bombay, Powai, Maharashtra 400076 India}

\begin{abstract}
We propose a novel origin for the disparity between quark and lepton flavor mixing based on the refractive nature of neutrino masses. We postulate that the fundamental mixing in both the quark and lepton sectors is CKM-like, together with tiny vacuum neutrino masses, while the observed PMNS mixing matrix emerges dynamically from coherent forward scattering of neutrinos on an ultralight dark matter background. The resulting in-medium Hamiltonian rotates CKM mixing angles into large effective lepton mixings, naturally realizing quark--lepton complementarity without invoking new flavor symmetries. This framework links neutrino mass generation, flavor mixing, and dark matter, and predicts environment-dependent neutrino oscillation effects testable in current and future experiments.
\end{abstract}

\maketitle

%%%%%%%%%%%%%%%%%%%%%%%%%%%%%%%%%%%%%%%%%%%%%%%%%%%%%%%%%%%%%%%%%%%%%%
%\noindent\textit{\textbf{Introduction}}---
%%%%%%%%%%%%%%%%%%%%%%%%%%%%%%%%%%%%%%%%%%%%%%%%%%%%%%%%%%%%%%%%%%%%%%
The origin of fermion flavor mixing remains one of the central unresolved problems of the Standard Model (SM). While quark mixing angles encoded in the Cabibbo--Kobayashi--Maskawa (CKM) matrix are small and hierarchical, lepton mixing angles appearing in the Pontecorvo--Maki--Nakagawa--Sakata (PMNS) matrix are strikingly large \cite{ParticleDataGroup:2024cfk,Esteban:2024eli}. This qualitative difference persists despite the parallel gauge structure and Yukawa origin of quark and lepton masses, suggesting that neutrino flavor may be governed by physics qualitatively distinct from that of charged fermions.

An intriguing empirical hint toward a unified description is provided by quark--lepton complementarity (QLC)~\cite{Smirnov:2004ju,Minakata:2004xt,Raidal:2004iw,Minakata:2005rf}, exemplified by the relation 
$\theta_{12}^{\rm PMNS} + \theta_{C} \simeq 45^\circ$,
which links the solar neutrino mixing angle, $\theta_{12}^{\rm PMNS}$, to the Cabibbo angle, $\theta_{C}$. This observation has motivated a wide class of models in which quark and lepton mixings originate from a common structure at high energies, often embedded in grand unification or discrete flavor symmetries~\cite{Raidal:2004iw,Frampton:2004vw,Ferrandis:2004vp,Kang:2005as,Antusch:2005ca,Schmidt:2006rb,Hochmuth:2006xn,Plentinger:2007px,Altarelli:2009gn}. In most such constructions, however, the PMNS matrix is assumed to be fundamental, with large lepton mixing angles imposed at the Lagrangian level. The physical origin of this largeness, and its sharp contrast with the quark sector, remains unexplained.

\begin{figure}[t!]
    \centering
    \includegraphics[width=0.9\linewidth]{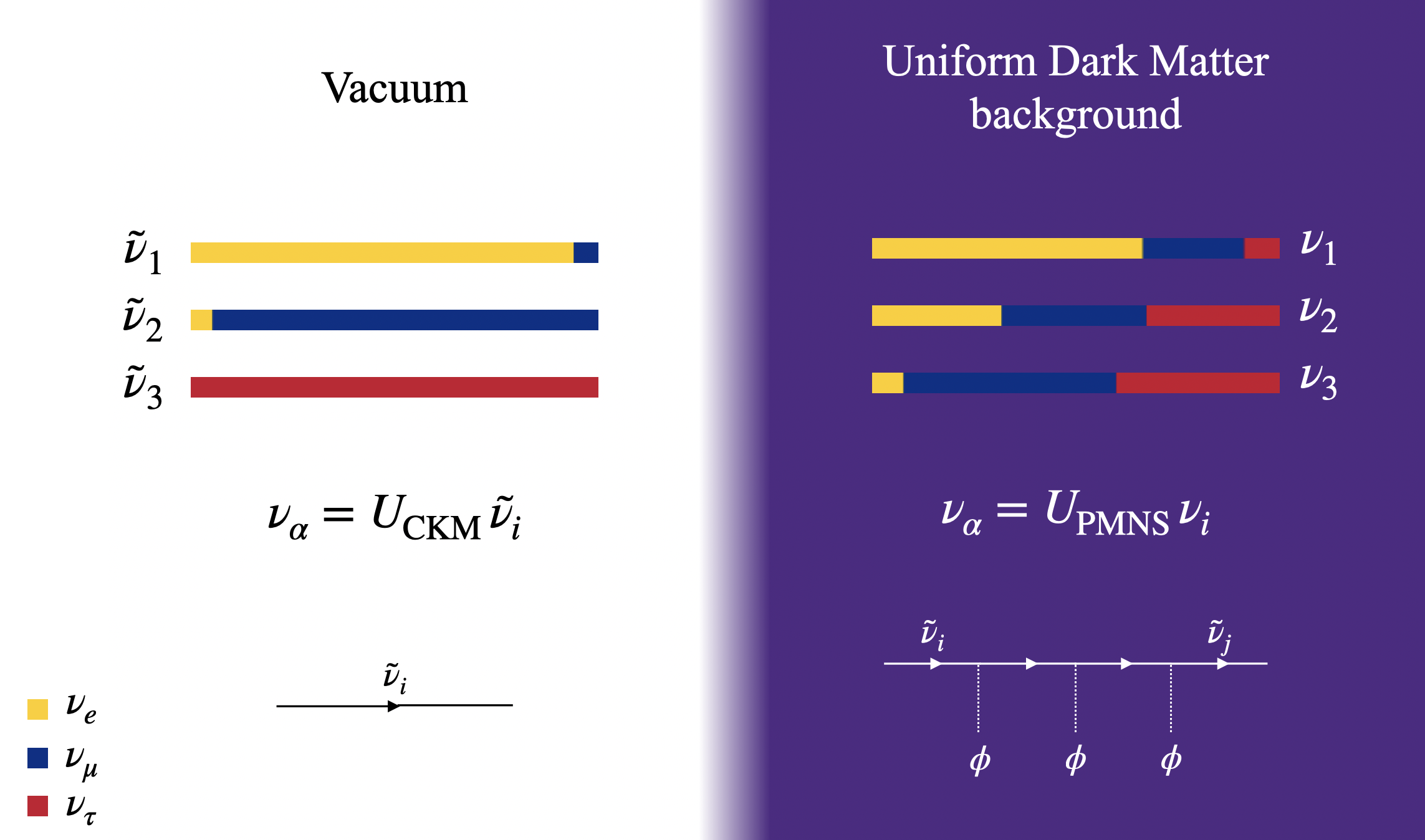}
    \caption{Illustration of our framework. In vacuum (left), neutrino mixing is CKM-like represented by the flavor content of the true mass eigenstates $\tilde\nu_i$. Meanwhile, in a uniform dark background experiments would measure a PMNS-like mixing summarized in the flavor composition of eigenstates in the dark-matter $\nu_i$.} \label{fig:illust}
\end{figure}

In this Letter, we explore a qualitatively different possibility in which the fundamental lepton mixing is CKM-like, while the observed PMNS matrix emerges dynamically as an environmental effect. Neutrinos are assumed to possess tiny vacuum masses with quark-like mixing, reflecting a unified quark--lepton flavor
structure at high energies. As illustrated in Fig.~\ref{fig:illust}, coherent
forward scattering on an ultralight dark matter (ULDM) background induces refractive contributions to the effective Hamiltonian during propagation, analogous in spirit to the Mikheyev--Smirnov--Wolfenstein (MSW) effect \cite{Mikheev:1986wj,Wolfenstein:1977ue}. When these contributions dominate over the vacuum masses, the effective mixing
matrix is rotated away from its CKM form, naturally reproducing the observed
PMNS pattern, including large solar and atmospheric mixing angles.

Here, quark-lepton complementarity arises dynamically, rather than being imposed as a symmetry of the fundamental theory. The largeness of lepton mixing angles is not a property of the underlying Yukawa sector, but a consequence of neutrino propagation in a dark-sector background. This sharply distinguishes our scenario from conventional flavor models and leads to distinctive phenomenological implications. In particular, the effective mixing parameters inferred from oscillation experiments may depend subtly on energy, baseline, or the local dark matter environment, offering new avenues to test the origin of neutrino flavor.

The possibility that neutrino masses are dominantly refractive has been developed in recent works~\cite{Ge:2019tdi,Choi:2019zxy,Ge:2020xkm,Ge:2020ffj,Choi:2020ydp,Smirnov:2021zgn,Chun:2021ief,Sen:2023uga,Ge:2024ftz,Perez-Gonzalez:2025qjh,Pompa:2025lbf,Chattopadhyay:2025ccy}, where it was shown that coherent forward scattering on ULDM can generate effective mass-squared terms with the same energy dependence as conventional vacuum masses at experimentally relevant energies, while allowing for nontrivial energy, time, or density dependence outside this regime. Since oscillation experiments probe only mass-squared differences, such refractive contributions can reproduce all observed oscillation data while evading cosmological bounds~\cite{DESI:2025zgx} and providing new observational handles~\cite{Berlin:2016woy,Brdar:2017kbt,Capozzi:2018bps,Dev:2020kgz,Losada:2021bxx,Huang:2022wmz,Dev:2022bae,Davoudiasl:2023uiq,Lopes:2023vxn,Martinez-Mirave:2024dmw,Goertz:2024gzw,Sahu:2025vyy,Sen:2024pgb,Chattopadhyay:2025ccy}.

This framework establishes a direct conceptual link between three longstanding puzzles: the origin of neutrino mass, the disparity between quark and lepton mixing, and the nature of dark matter. In the remainder of this Letter, we present a minimal realization of this idea, outline the conditions under which CKM-like vacuum mixing is rotated into PMNS-like mixing by refractive effects, and discuss the resulting phenomenological signatures in solar, atmospheric, reactor, and long-baseline neutrino experiments.

%%%%%%%%%%%%%%%%%%%%%%%%%%%%%%%%%%%%%%%%%%%%%%%%%%%%%%%%%%%%%%%%%%%%%%
\noindent\textit{\textbf{PMNS matrix from refractive neutrino masses}}--- 
%%%%%%%%%%%%%%%%%%%%%%%%%%%%%%%%%%%%%%%%%%%%%%%%%%%%%%%%%%%%%%%%%%%%%%
The relevant section of the Lagrangian capturing the interaction of neutrinos with ULDM, $\phi$, through the  mediator $\chi$, is
\begin{equation}
 \mathcal{L} \supset \sum_{\substack{\alpha=e,\mu,\tau \\ k=1,2}} \biggl((M_{\rm vac})_{\alpha\beta}\,\bar{\nu}_{\alpha}\nu_{\beta}  + g_{\alpha k}\bar{\nu}_{\alpha}\chi_{k}\,\phi^\star +  m_{\chi_k}\,\bar{\chi}_{k}\chi_{k} \biggr)\,,
\end{equation}
while $g$ denotes the interaction couplings and $m_\chi$ are the masses of the mediators. The major difference with~\cite{Sen:2023uga} is that neutrinos are assumed to have tiny masses, governed by the mass matrix, $M_{\rm vac} = [V_{\text{CKM}}]\cdot\,{\rm diag} (m_1,m_2,m_3)\cdot\,[V_{\text{CKM}}^\dagger]\,$, appropriately rotated by the CKM mixing matrix~\footnote{We note that the vacuum masses could be diagonalized by a different matrix in general. We consider as an example the CKM case while keeping in mind that our framework is applicable to other case.}.

Neutrinos get the major contribution to their mass through forward scattering on the ULDM background. The effective potential depends on the ULDM density and exhibits a resonance
at energy $E_R = m_\chi^2/(2m_\phi)$, where $m_\phi$ is the ULDM mass. For a single neutrino flavor, for energies above $E_R$, the potential reduces to
$V \simeq m_{\rm dark}^2/(2E_\nu)$, with $m_{\rm dark}^2 = g^2\rho_\phi/m_\phi^2$, governed by the local DM density $\rho_\phi$, 
reproducing the standard energy dependence required by neutrino oscillations \cite{Sen:2023uga}.  The expectation value of the coherent ULDM can be parameterised as $\langle \phi \rangle^*_{\rm coh}=F e^{i \Phi}$~\cite{Sen:2023uga}, where $F$ is the amplitude and $\Phi$ is a phase dependent on the ULDM mass.

The crucial point of our scenario is that, as long as the vacuum contribution is much smaller than the refractive contribution, $M_{\rm vac} \ll F[g]$, one can generate the PMNS mixing matrix by choosing couplings such that
\begin{equation}
     F\, [g] = [U_{\text{PMNS}}] \cdot [\Delta M_{\rm D}] \cdot [W^T]\,,\label{eq:SVD}
\end{equation}
where $U_{\rm PMNS}$ is the usual $3\times 3$ PMNS mixing matrix,  
\begin{equation}\label{eq:DMD}
     \Delta M_{\rm D} \equiv \begin{pmatrix}
        0 & 0\\
        \sqrt{\Delta m^2_\text{sol}} & 0\\
        0 & \sqrt{\Delta m^2_\text{atm}}
    \end{pmatrix}\, ,
\end{equation}
with $\Delta m^2_\text{sol}$ and $\Delta m^2_\text{atm}$ the observed solar and atmospheric mass-squared differences, respectively, and $W$ is any arbitrary $2\times2$ real orthogonal matrix required for diagonalisation. Note that $\Delta M_{\rm D}\cdot \Delta M_{\rm D}^T =  {\rm diag}(0, \Delta m^2_\text{sol}, \Delta m^2_\text{atm}) \equiv \Delta M_{\rm D}^2 $.

After a rotation of $\lvert \chi_{k}\rangle \to e^{-i \Phi}\lvert \chi_k\rangle$, the resulting Hamiltonian in the basis $\lvert \widetilde{\nu}_\alpha \rangle = \left\{\lvert\nu_e\rangle,\lvert\nu_\mu\rangle,\lvert\nu_\tau\rangle,e^{-i\Phi}\lvert\chi_1\rangle,e^{-i\Phi}\lvert\chi_2\rangle\right\}$ can be written as a $5\times5$ mass-matrix as
\begin{widetext}
\begin{align}\label{eq:finHam}
    \widetilde{\mathbb{H}} =& \frac{1}{2E} \begin{pmatrix}
        U_{\rm PMNS} \cdot \Delta M_{\rm D}^2 \cdot U_{\rm PMNS}^\dagger &\qquad  U_{\rm PMNS} \cdot \Delta M_{\rm D} \cdot \tilde{m}_{\chi_k} \\
        \tilde{m}_{\chi_k}^T \cdot \Delta M_{\rm D}^T \cdot U_{\rm PMNS}^\dagger &\qquad [\Delta M_D^2]_{2\times 2} + [m_{\chi_k}^2+ 2E\dot{\Phi}]_D
        \end{pmatrix} \notag\\
        &+ \frac{1}{2E} \begin{pmatrix}
            M_{\rm vac} \cdot M_{\rm vac}^{\dagger} &\ & e^{2i\Phi}\, M_{\rm vac} \cdot U_{\rm PMNS}^* \cdot \Delta M_{\rm D} \cdot W^\dagger \\
            e^{-2i\Phi}\, W\cdot \Delta M_{\rm D}^T \cdot U_{\rm PMNS}^T  \cdot M_{\rm vac}^{\dagger} &\ & [0]_{2\times2}
        \end{pmatrix}\,,
\end{align}
\end{widetext}
where $\tilde{m}_{\chi_k}$ is a $2\times 2$ matrix related to the masses of the $\chi_{1,2}$ states, $[\Delta M_D^2]_{2\times 2} = \Delta M_{\rm D}^T\cdot \Delta M_{\rm D} =  {\rm diag}(\Delta m^2_\text{sol}, \Delta m^2_\text{atm})$ refers to the bottom-right $2\times2$ block of $\Delta M_{\rm D}^2$ (see Supplemental Material for further details).

The entire Hamiltonian $\widetilde{\mathbb{H}} $ can be diagonalised by a $5\times 5$ unitary matrix, which contains the PMNS matrix as a submatrix diagonalising the $3\times 3$ active block of  $\widetilde{\mathbb{H}} $.
Thus, a CKM-like mixing in vacuum may be observed as a PMNS-like mixing due to the effects of background DM. 

Eq.\,\ref{eq:finHam} denotes the full Hamiltonian for propagation of neutrinos in the ULDM background. Throughout this work, we have chosen the vacuum mass to be $m_1=0\, \text{and}\, m_3=2\, m_2 \equiv 2\, \delta m$. We note that constraints from oscillation experiments will be modified if we change this benchmark spectrum. We will consider more generic scenarios elsewhere.

The results from different neutrino experiments have to be interpreted in the light of this scenario in order to check the viability of this idea. We proceed with this in the following sections. Unless specified otherwise, we have used the best fit values for the observed active mixing angles, i.e. $\theta_{12} = 33.76^{\circ},\, \theta_{23} = 43.27^{\circ},\, \theta_{31} = 8.62^{\circ}, \delta_{\rm CP} = 207^{\circ}$ and mass-squared differences, i.e. $\Delta m_{\rm sol}^2 = 7.537 \times 10^{-5}\, {\rm eV}^2$ and $\Delta m_{\rm atm}^2 = 2.521 \times 10^{-3}\, {\rm eV}^2$ \cite{Esteban:2024eli}. Moreover, we had chosen the rotation angle of $[W]$ to be $25^{\circ}$.  

%%%%%%%%%%%%%%%%%%%%%%%%%%%%%%%%%%%%%%%%%%%%%%%%%%%%%%%%%%%%%%%%%%%%%%
\noindent\textit{\textbf{Solar Neutrinos}}---
%%%%%%%%%%%%%%%%%%%%%%%%%%%%%%%%%%%%%%%%%%%%%%%%%%%%%%%%%%%%%%%%%%%%%%
%============================================%
\begin{figure}[!t]
		\centering
		\includegraphics[width=0.9\linewidth]{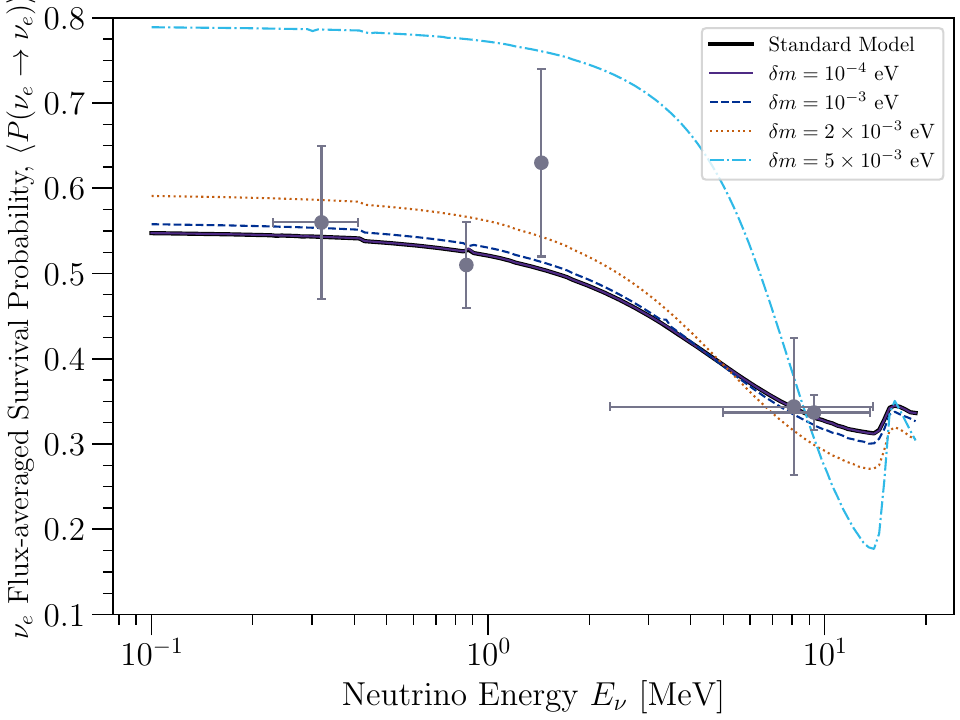}
		\caption{Solar neutrino flux-averaged survival probability in the Standard scenario (black) and in our model for $\delta m = 10^{-4}~{\rm eV}$ (purple), $10^{-3}~{\rm eV}$ (dashed blue), $2\times 10^{-3}~{\rm eV}$ (dotted orange), and $5\times 10^{-3}~{\rm eV}$ (dot-dashed light blue). The gray data indicate the different measurements from solar neutrino experiments.} \label{fig:solar_nu}
\end{figure} 
%============================================%
Solar neutrinos are produced via different reaction channels inside the Sun
%with each channel having its characteristic flux as well as radial production distributions. 
Fig.\,\ref{fig:solar_nu} shows the flux-averaged survival probability $P_{ee}$ of solar neutrinos in the presence and absence of the ULDM background. We find that even for $\delta m = 2 \times 10^{-3}$ eV ($\delta m^2 = 4 \times 10^{-6}~{\rm eV^2}$), there can be significant changes in the flux-averaged probability spectrum. The upturn of the spectrum becomes steeper compared to the usual MSW solution. This happens because, $\theta_{12}^{\text{CKM}} < \theta_{12}^{\text{PMNS}}$ which results in $P_{ee}^{\text{low,\,refr.}} > P_{ee}^{\text{low,\,MSW}}$ and $P_{ee}^{\text{high,\,refr.}} < P_{ee}^{\text{high,\,MSW}}$, where $P_{ee}^{\text{low}} = \cos^4 \theta_{12} + \sin^4 \theta_{12}$ is the survival probability for low energies ($\approx 0.1\, \text{MeV}$) and $P_{ee}^{\text{high}} = \sin^2 \theta_{12}$ is the survival probability for high energies ($\approx 10\, \text{MeV}$). Thus, the data from solar neutrino experiments disfavor $\delta m \gtrsim 10^{-3}$ eV. 
%============================================%
\begin{figure}[t!]
		\centering
		\includegraphics[width=\linewidth]{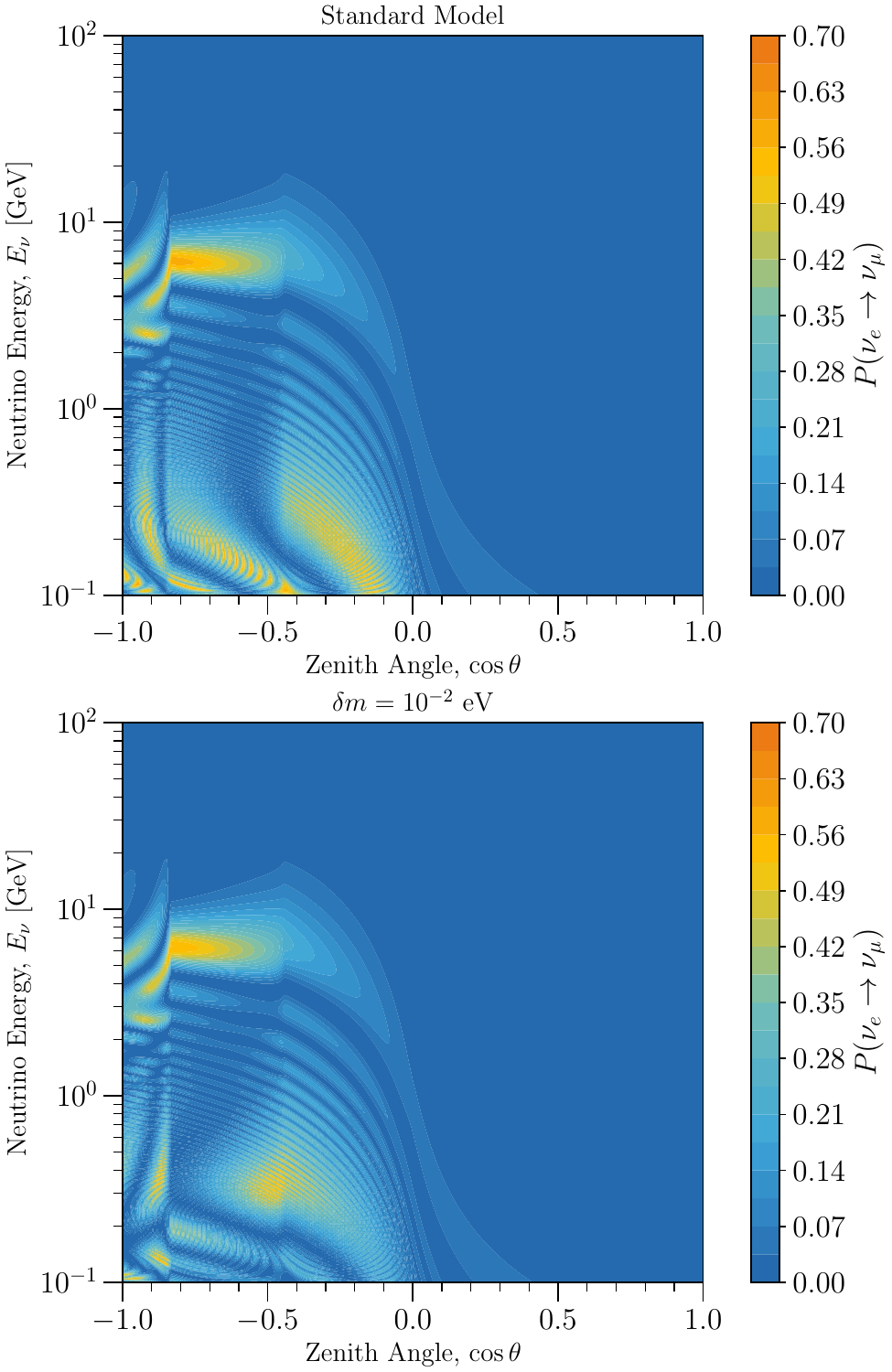}
		\caption{Atmospheric neutrino oscillogram for the $\nu_e\to\nu_\mu$ channel in terms of neutrino energy and zenith angle. We present the standard case (left) together with the oscillation pattern in our framework with $\delta m=10^{-2}~{\rm eV}$ (right).} \label{fig:atm_nu}
\end{figure} 
%============================================%

%%%%%%%%%%%%%%%%%%%%%%%%%%%%%%%%%%%%%%%%%%%%%%%%%%%%%%%%%%%%%%%%%%%%%%
\noindent\textit{\textbf{Atmospheric Neutrinos}}---
%%%%%%%%%%%%%%%%%%%%%%%%%%%%%%%%%%%%%%%%%%%%%%%%%%%%%%%%%%%%%%%%%%%%%%
Atmospheric neutrinos span a large energy range, from $\sim 100$ MeV to $~{\cal O}(1)$ TeV.
Since these neutrinos reach detectors from different directions, oscillations of atmospheric neutrinos are quite rich, given the internal structure of Earth.
Specifically, neutrinos could develop parametric and MSW resonances depending on their energies and on whether they transverse different matter layers~\cite{Liu:1998nb,Akhmedov:1998ui,Chizhov:1998ug,Akhmedov:1998xq,Chizhov:1999he,Akhmedov:2006hb,Kelly:2021jfs}.
As an example, we present in Fig.~\ref{fig:atm_nu} (top) the standard $P(\nu_e\to \nu_\mu)$ oscillation probability for values of neutrino energy and zenith angle $\theta$, which parametrizes the direction of the incoming neutrino.
We observe the parametric resonances developing for values of $\cos\theta\lesssim -0.85$ while for $-0.85\lesssim \cos\theta \lesssim -0.45$ MSW resonances appear for $E_\nu\sim 6~{\rm GeV}$ for oscillations driven by $\theta_{13}$ and $\Delta m_{31}^2$.
Now, if we consider our framework for a benchmark value of $\delta m = 10^{-2}~{\rm eV}$, we observe that the whole oscillation pattern is significantly modified. 
Specifically, the parametric and MSW resonances are shifted for different values of neutrino energy due to the significant contribution of the true vacuum term, particularly for the sub-GeV region, where oscillations are dominated by solar parameters $\theta_{12},\, \Delta m_{21}^2$. 
Similar behavior is observed in other oscillation channels.
Therefore, the precise measurement of atmospheric neutrinos, specially in the sub-GeV region, see e.g.~Ref.~\cite{Kelly:2019itm}, could put additional constraints in our framework.

%%%%%%%%%%%%%%%%%%%%%%%%%%%%%%%%%%%%%%%%%%%%%%%%%%%%%%%%%%%%%%%%%%%%%%
\noindent\textit{\textbf{Long baseline experiments}}---
%%%%%%%%%%%%%%%%%%%%%%%%%%%%%%%%%%%%%%%%%%%%%%%%%%%%%%%%%%%%%%%%%%%%%%
%============================================%
\begin{figure*}[t!]
		\centering
		\includegraphics[width=0.95\linewidth]{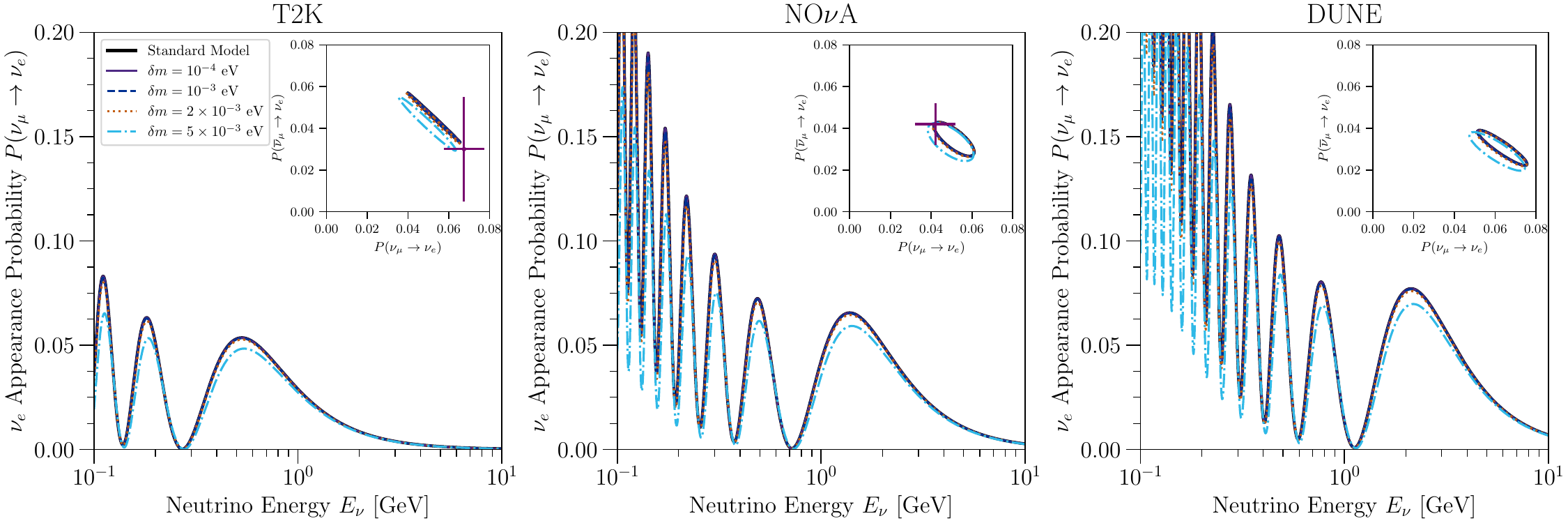}
		\caption{$\nu_e$ appearance probability $P(\nu_\mu\to\nu_e)$ as function of neutrino energy for different long baseline experiments, T2K~\cite{T2K:2023smv} (left), NO$\nu$A~\cite{NOvA:2021nfi} (middle) and DUNE~\cite{DUNE:2020jqi} (right). We present the standard oscillation (black) and the probability obtained in our framework for or $\delta m = 10^{-4}~{\rm eV}$ (purple), $10^{-3}~{\rm eV}$ (dashed blue), $2\times 10^{-3}~{\rm eV}$ (dotted orange), and $5\times 10^{-3}~{\rm eV}$ (dot-dashed light blue). The inset indicate the biprobabilities $P(\bar\nu_\mu\to\bar\nu_e)$ vs $P(\nu_\mu\to\nu_e)$ for the same experiments together with current measurements in T2K and NO$\nu$A.} \label{fig:longbase}
\end{figure*} 
%============================================%
We consider next the effects of our framework in long baseline experiments, such as T2K~\cite{T2K:2023smv}, NO$\nu$A~\cite{NOvA:2021nfi} and DUNE~\cite{DUNE:2020jqi}.
In these experiments, a $\sim 90\%$ pure $\nu_\mu$ flux with energies of $\sim{\cal O}$(GeV) is produced from focused mesons emitted after a proton beam impinged some target material.
Such a flux is directed towards a neutrino detector at a distance of ${\cal O}(100~{\rm km})$.
These facilities simultaneously search for $\nu_\mu$ disappearance and $\nu_e$ appearance to measure the $\Delta m_{31}^2, \sin^2\theta_{23}$ and $\delta_{\rm CP}$ parameters, respectively.
We present in Fig.~\ref{fig:longbase} the appearance probability of electron neutrinos in constant matter $P(\nu_\mu\to\nu_e)$ as function of neutrino energy for T2K (left), NO$\nu$A (middle) and DUNE (right) for the same parametrization of the true vacuum masses as before. 
For completeness, we also present as inset figures biprobability plots for the same experiments, computed for neutrino energies of $0.6$ GeV (T2K), $2.1$ GeV (NO$\nu$A) and $3$ GeV (DUNE), together with the measured appearance neutrino and antineutrino probabilities for T2K and NO$\nu$A, similar to Ref.~\cite{Kelly:2020fkv}.
We observe that our scenario reproduces standard oscillations for values $\delta m \lesssim 2\times 10^{-3}~{\rm eV}$, while only minor modifications appear for $\delta m = 5\times 10^{-3}~{\rm eV}$.
We have also verified that the muon neutrino disappearance probability $P(\nu_\mu\to\nu_\mu)$ is also marginally affected in our framework.

%%%%%%%%%%%%%%%%%%%%%%%%%%%%%%%%%%%%%%%%%%%%%%%%%%%%%%%%%%%%%%%%%%%%%%
\noindent\textit{\textbf{Reactor experiments}}---
%%%%%%%%%%%%%%%%%%%%%%%%%%%%%%%%%%%%%%%%%%%%%%%%%%%%%%%%%%%%%%%%%%%%%%
%============================================%
\begin{figure}[t!]
		\centering
		\includegraphics[width=0.9\linewidth]{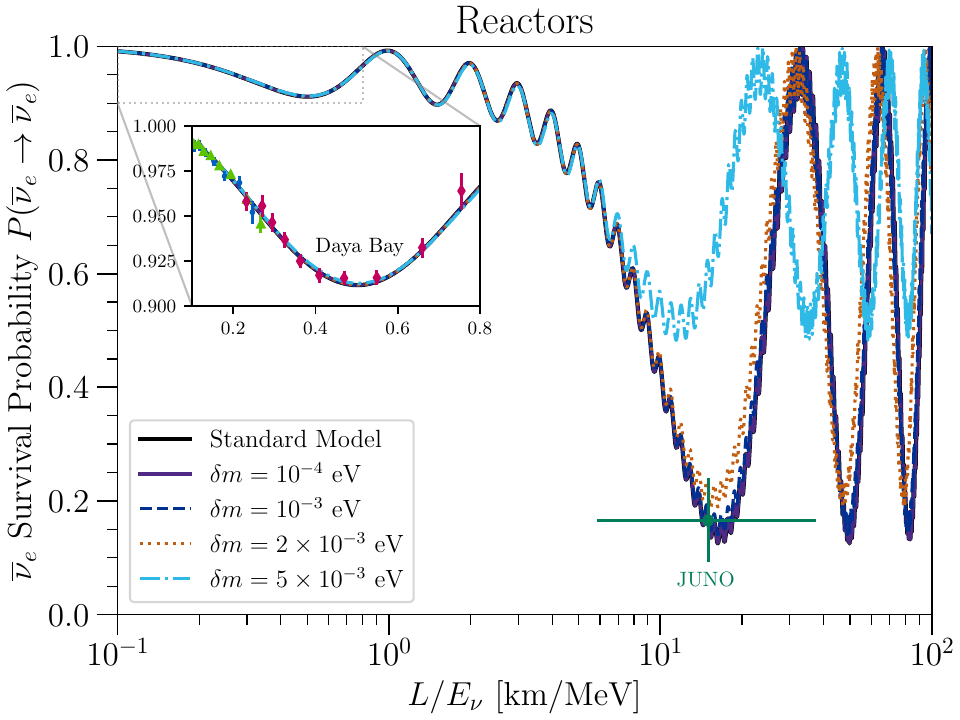}
		\caption{Electron antineutrino survival probability $P(\bar\nu_e\to\bar\nu_e)$ as function of $L/E_{\nu}$ for standard oscillations (black) and our framework assuming $\delta m = 10^{-4}~{\rm eV}$ (purple), $10^{-3}~{\rm eV}$ (dashed blue), $2\times 10^{-3}~{\rm eV}$ (dotted orange), and $5\times 10^{-3}~{\rm eV}$ (dot-dashed light blue). The inset shows the Daya Bay region-of-interest together with the data points from Ref.~\cite{DayaBay:2022orm}. We also present the recent JUNO result~\cite{JUNO:2025gmd} as a green point.} \label{fig:reactors}
\end{figure} 
%============================================%
In a reactor, fission of $^{235}$U and subsequent decay processes generate an intense electron antineutrino flux $\gtrsim 10^{21}\, \bar\nu_e/{\rm s}$ with energies of $\sim{\cal O}$(MeV).
Given their energies, these antineutrinos have been used to measure the 1-3 mixing $\theta_{13}$ and the effective $\Delta m_{ee}^2$ mass splitting~\cite{Nunokawa:2005nx} and the solar parameters $\Delta m_{21}^2,\ \sin^2\theta_{12}$ in short ${\cal O}(1~{\rm km})$ and medium distance ${\cal O}(50~{\rm km})$ experiments, respectively. 
Among several experiments, the most precise determination of $\theta_{13}$ with a precision of $\sim 1\%$ has been achieved by the Daya Bay experiment~\cite{DayaBay:2022orm}, while, for the solar sector, the JUNO experiment has recently announced a high-precision measurement with only 59.1 days of data taking~\cite{JUNO:2025gmd}.
Therefore, we analyze the effects of our framework on reactor antineutrino oscillations.
We present the $\bar\nu_e$ survival probability $P(\bar\nu_e\to\bar\nu_e)$ as function of $L/E_\nu$ in Fig.~\ref{fig:reactors} for standard oscillations and in our scenario.
We also show the Daya Bay data in the inset and the recent JUNO measurement, obtained by varying the solar parameters in the $3\sigma$ range as given in Ref.~\cite{JUNO:2025gmd}. 
For the chosen benchmarks, we observe that the short baseline oscillations as measured by Daya Bay are not modified by the vacuum masses and mixings.
However, for $L/E_\nu\gtrsim 7~{\rm km/MeV}$ and values $\delta m\gtrsim 2\times 10^{-3}~{\rm eV}$ the true vacuum term starts interfering significantly with the dark matter term, modifying the oscillation pattern so that such parameters become disfavored by the JUNO measurement.
Although a full analysis of both Daya Bay and JUNO data is beyond the scope of this Letter and it is left for future work, we expect that these data set will provide stringent limits in our framework.

%%%%%%%%%%%%%%%%%%%%%%%%%%%%%%%%%%%%%%%%%%%%%%%%%%%%%%%%%%%%%%%%%%%%%%
\noindent\textit{\textbf{Cosmology and astrophysics}}---
%%%%%%%%%%%%%%%%%%%%%%%%%%%%%%%%%%%%%%%%%%%%%%%%%%%%%%%%%%%%%%%%%%%%%%
The impact on the early Universe cosmology has already been discussed in~\cite{Sen:2024pgb}. Below the resonance energy, $E_R$, the refractive mass decreases with energy in a manner controlled by the DM asymmetry, rendering it ineffective as a conventional mass term. As a result, refractive neutrino masses are negligible during structure formation, effectively leaving relic neutrinos massless and allowing this scenario to remain consistent with stringent cosmological limits on the sum of neutrino masses. This yields a lower limit on $E_R \gtrsim 11\,{\rm eV}$. Furthermore, Ref.~\cite{Cheek:2025kks} used KamLAND data to place bounds disfavouring DM-induced time-modulated effects as the dominant source of neutrino
mass, assuming rapid decoherence from $\mathcal{O}(1)$ DM fluctuations.
However, uncertainties in the long-term coherence of the DM field and in
the modeling of realistic galactic substructure leave room for alternative
interpretations, and since neutrinos in our framework have nonzero vacuum masses
with dark matter providing only a subdominant contribution, these bounds do not
directly apply. 

Astrophysical neutrinos place strong constraints on refractive neutrino mass scenarios. Neutrinos from SN1987A and TeV--PeV neutrinos observed by IceCube must traverse galactic and cosmological DM backgrounds without significant attenuation, thereby bounding
neutrino - DM interactions over a wide energy range. At the same time, refractive
masses modify neutrino propagation through time-of-flight delays that depend on the
integrated DM density. For a galactic core-collapse supernova, such delays can
reach $10^{-2}-1\,\mathrm{s}$ for oscillation-scale effective masses, within the sensitivity
of next-generation detectors such as DUNE. The consistency of flux observations together
with the prospect of observable time delays tightly constrains and sharply tests this
framework using astrophysical data alone.

%%%%%%%%%%%%%%%%%%%%%%%%%%%%%%%%%%%%%%%%%%%%%%%%%%%%%%%%%%%%%%%%%%%%%%
\noindent\textit{\textbf{Final thoughts}}---
%%%%%%%%%%%%%%%%%%%%%%%%%%%%%%%%%%%%%%%%%%%%%%%%%%%%%%%%%%%%%%%%%%%%%%
We have presented a new paradigm for the origin of lepton flavor mixing in which the striking disparity between quark and lepton mixing angles is not fundamental but emergent. Neutrinos possess only tiny vacuum masses with quark-like mixing at the most basic level. Coherent forward scattering on an ultralight dark matter background generates refractive contributions to the propagation Hamiltonian, dynamically rotating this small fundamental mixing into the large effective lepton mixing observed in experiments. This complementarity between quark and leptonic mixing thus arises naturally, without new flavor symmetries, textures, or tuning, and directly ties neutrino flavor to the dark sector.

Most importantly, this scenario is experimentally testable. Because refractive effects depend on neutrino energy, baseline, and the properties of the surrounding dark matter, different oscillation experiments need not infer identical effective mixing parameters. Precision measurements from solar, atmospheric, reactor, and long-baseline neutrino experiments therefore offer a unique opportunity to probe this mechanism. Upcoming and next-generation facilities, with improved sensitivity to subtle deviations from standard oscillation expectations, can directly test whether the large lepton mixing angles observed in nature are a manifestation of neutrino refraction in the dark sector.

%%%%%%%%%%%%%%%%%%%%%%%%%%%%%%%%%%%%%%%%%%%%%%%%%%%%%%%%%%%%%%%%%%%%%%
\begin{acknowledgments}

We thank Alexei Smirnov, André de Gouvêa and Pedro Machado for a careful reading of the manuscript.
We would also like to thank Amol Dighe, Enrique Fernández-Martínez and Michele Maltoni for the useful discussions.
YFPG acknowledges financial support by the Consolidaci\'on Investigadora grant CNS2023-144536 from the Spanish Ministerio de Ciencia e Innovaci\'on (MCIN) and by the Spanish Research Agency (Agencia Estatal de Investigaci\'on) through the grant IFT Centro de Excelencia Severo Ochoa No CEX2020-001007-S. MS acknowledges support from the Early Career Research Grant by Anusandhan National Research Foundation (project number ANRF/ECRG/2024/000522/PMS). MS also acknowledges support from the IoE-funded Seed Funding for Collaboration and Partnership Projects - Phase IV SCPP grant (RD/0524-IOE00I0-012) by IIT Bombay. MS and YP would also like to thank the Galileo Galilei Institute for Theoretical Physics for the hospitality and the INFN for partial support during the beginning of this work. 

\end{acknowledgments}
%%%%%%%%%%%%%%%%%%%%%%%%%%%%%%%%%%%%%%%%%%%%%%%%%%%%%%%%%%%%%%%%%%%%%%

\bibliography{biblio.bib}

@article{ParticleDataGroup:2024cfk,
    author = "Navas, S. and others",
    collaboration = "Particle Data Group",
    title = "{Review of particle physics}",
    doi = "10.1103/PhysRevD.110.030001",
    journal = "Phys. Rev. D",
    volume = "110",
    number = "3",
    pages = "030001",
    year = "2024"
}

@article{Esteban:2024eli,
    author = "Esteban, Ivan and Gonzalez-Garcia, M. C. and Maltoni, Michele and Martinez-Soler, Ivan and Pinheiro, Jo{\~a}o Paulo and Schwetz, Thomas",
    title = "{NuFit-6.0: updated global analysis of three-flavor neutrino oscillations}",
    eprint = "2410.05380",
    archivePrefix = "arXiv",
    primaryClass = "hep-ph",
    reportNumber = "IFT-UAM/CSIC-24-140, YITP-SB-2024-24, IPPP/24/64, IPPP/24/64, IFT-UAM/CSIC-24-140, YITP-SB-2024-24",
    doi = "10.1007/JHEP12(2024)216",
    journal = "JHEP",
    volume = "12",
    pages = "216",
    year = "2024"
}

@article{Sen:2023uga,
    author = "Sen, Manibrata and Smirnov, Alexei Y.",
    title = "{Refractive neutrino masses, ultralight dark matter and cosmology}",
    eprint = "2306.15718",
    archivePrefix = "arXiv",
    primaryClass = "hep-ph",
    doi = "10.1088/1475-7516/2024/01/040",
    journal = "JCAP",
    volume = "01",
    pages = "040",
    year = "2024"
}

@article{JUNO:2025gmd,
    author = "Abusleme, Angel and others",
    collaboration = "JUNO",
    title = "{First measurement of reactor neutrino oscillations at JUNO}",
    eprint = "2511.14593",
    archivePrefix = "arXiv",
    primaryClass = "hep-ex",
    journal = "",
    month = "11",
    year = "2025"
}

@article{Sen:2024pgb,
    author = "Sen, Manibrata and Smirnov, Alexei Y.",
    title = "{Neutrinos with refractive masses and the DESI baryon acoustic oscillation results}",
    eprint = "2407.02462",
    archivePrefix = "arXiv",
    primaryClass = "hep-ph",
    doi = "10.1103/d9hh-b3r9",
    journal = "Phys. Rev. D",
    volume = "111",
    number = "10",
    pages = "103048",
    year = "2025"
}

@article{Cheek:2025kks,
    author = "Cheek, Andrew and Visinelli, Luca and Zhang, Hong-Yi",
    title = "{Testing the Dark Origin of Neutrino Masses with Oscillation Experiments}",
    eprint = "2503.08439",
    archivePrefix = "arXiv",
    primaryClass = "hep-ph",
    doi = "10.1103/wyns-m4y5",
    journal = "Phys. Rev. Lett.",
    volume = "135",
    number = "3",
    pages = "031801",
    year = "2025"
}

@article{Mikheev:1986wj,
    author = "Mikheev, S. P. and Smirnov, A. Yu.",
    title = "{Resonant amplification of neutrino oscillations in matter and solar neutrino spectroscopy}",
    doi = "10.1007/BF02508049",
    journal = "Nuovo Cim. C",
    volume = "9",
    pages = "17--26",
    year = "1986"
}

@article{Wolfenstein:1977ue,
    author = "Wolfenstein, L.",
    title = "{Neutrino Oscillations in Matter}",
    reportNumber = "COO-3066-102",
    doi = "10.1103/PhysRevD.17.2369",
    journal = "Phys. Rev. D",
    volume = "17",
    pages = "2369--2374",
    year = "1978"
}

@article{Ge:2019tdi,
    author = "Ge, Shao-Feng and Murayama, Hitoshi",
    title = "{Apparent CPT Violation in Neutrino Oscillation from Dark Non-Standard Interactions}",
    eprint = "1904.02518",
    archivePrefix = "arXiv",
    primaryClass = "hep-ph",
    reportNumber = "IPMU19-0047",
    journal = "",
    month = "4",
    year = "2019"
}

@article{Ge:2020xkm,
    author = "Ge, Shao-Feng",
    title = "{The Leptonic CP Measurement and New Physics Alternatives}",
    doi = "10.22323/1.369.0108",
    journal = "PoS",
    volume = "NuFact2019",
    pages = "108",
    year = "2020"
}

@article{Ge:2020ffj,
    author = "Ge, Shao-Feng",
    editor = "Nakahata, Masayuki",
    title = "{New Physics with Scalar and Dark Non-Standard Interactions in Neutrino Oscillation}",
    doi = "10.1088/1742-6596/1468/1/012125",
    journal = "J. Phys. Conf. Ser.",
    volume = "1468",
    number = "1",
    pages = "012125",
    year = "2020"
}

@article{Choi:2019zxy,
    author = "Choi, Ki-Young and Chun, Eung Jin and Kim, Jongkuk",
    title = "{Neutrino Oscillations in Dark Matter}",
    eprint = "1909.10478",
    archivePrefix = "arXiv",
    primaryClass = "hep-ph",
    doi = "10.1016/j.dark.2020.100606",
    journal = "Phys. Dark Univ.",
    volume = "30",
    pages = "100606",
    year = "2020"
}

@article{Choi:2020ydp,
    author = "Choi, Ki-Young and Chun, Eung Jin and Kim, Jongkuk",
    title = "{Dispersion of neutrinos in a medium}",
    eprint = "2012.09474",
    archivePrefix = "arXiv",
    primaryClass = "hep-ph",
    reportNumber = "KIAS-P20072",
    journal = "",
    month = "12",
    year = "2020"
}

@article{Smirnov:2021zgn,
    author = "Smirnov, Alexei Y. and Valera, Victor B.",
    title = "{Resonance refraction and neutrino oscillations}",
    eprint = "2106.13829",
    archivePrefix = "arXiv",
    primaryClass = "hep-ph",
    doi = "10.1007/JHEP09(2021)177",
    journal = "JHEP",
    volume = "09",
    pages = "177",
    year = "2021"
}

@article{Chun:2021ief,
    author = "Chun, Eung Jin",
    title = "{Neutrino Transition in Dark Matter}",
    eprint = "2112.05057",
    archivePrefix = "arXiv",
    primaryClass = "hep-ph",
    reportNumber = "KIAS-P21056",
    journal = "",
    month = "12",
    year = "2021"
}

@article{Ge:2024ftz,
    author = "Ge, Shao-Feng and Kong, Chui-Fan and Smirnov, Alexei Y.",
    title = "{Testing the Origins of Neutrino Mass with Supernova-Neutrino Time Delay}",
    eprint = "2404.17352",
    archivePrefix = "arXiv",
    primaryClass = "hep-ph",
    doi = "10.1103/PhysRevLett.133.121802",
    journal = "Phys. Rev. Lett.",
    volume = "133",
    number = "12",
    pages = "121802",
    year = "2024"
}

@article{Perez-Gonzalez:2025qjh,
    author = "Perez-Gonzalez, Yuber F. and Sen, Manibrata",
    title = "{Dynamic Neutrino Mass Ordering and Its Imprint on the Diffuse Supernova Neutrino Background}",
    eprint = "2501.16412",
    archivePrefix = "arXiv",
    primaryClass = "hep-ph",
    reportNumber = "IFT-UAM/CSIC-25-4",
    journal = "",
    month = "1",
    year = "2025"
}

@article{Pompa:2025lbf,
    author = "Pompa, Federica and Sen, Manibrata",
    title = "{Shedding light on dark matter spikes through neutrino-dark matter interactions}",
    eprint = "2508.10983",
    archivePrefix = "arXiv",
    primaryClass = "hep-ph",
    journal = "",
    month = "8",
    year = "2025"
}

@article{Chattopadhyay:2025ccy,
    author = "Chattopadhyay, Susobhan and Dighe, Amol",
    title = "{Refractive neutrino masses in the solar DM halo: Can the dark-LMA solution be revived?}",
    eprint = "2511.19420",
    archivePrefix = "arXiv",
    primaryClass = "hep-ph",
    reportNumber = "TIFR/TH/25-20",
    journal = "",
    month = "11",
    year = "2025"
}

@article{Berlin:2016woy,
    author = "Berlin, Asher",
    title = "{Neutrino Oscillations as a Probe of Light Scalar Dark Matter}",
    eprint = "1608.01307",
    archivePrefix = "arXiv",
    primaryClass = "hep-ph",
    doi = "10.1103/PhysRevLett.117.231801",
    journal = "Phys. Rev. Lett.",
    volume = "117",
    number = "23",
    pages = "231801",
    year = "2016"
}

@article{Brdar:2017kbt,
    author = "Brdar, Vedran and Kopp, Joachim and Liu, Jia and Prass, Pascal and Wang, Xiao-Ping",
    title = "{Fuzzy dark matter and nonstandard neutrino interactions}",
    eprint = "1705.09455",
    archivePrefix = "arXiv",
    primaryClass = "hep-ph",
    reportNumber = "MITP-17-037",
    doi = "10.1103/PhysRevD.97.043001",
    journal = "Phys. Rev. D",
    volume = "97",
    number = "4",
    pages = "043001",
    year = "2018"
}

@article{Capozzi:2018bps,
    author = "Capozzi, Francesco and Shoemaker, Ian M. and Vecchi, Luca",
    title = "{Neutrino Oscillations in Dark Backgrounds}",
    eprint = "1804.05117",
    archivePrefix = "arXiv",
    primaryClass = "hep-ph",
    doi = "10.1088/1475-7516/2018/07/004",
    journal = "JCAP",
    volume = "07",
    pages = "004",
    year = "2018"
}

@article{Dev:2020kgz,
    author = "Dev, Abhish and Machado, Pedro A. N. and Mart{\'\i}nez-Mirav{\'e}, Pablo",
    title = "{Signatures of ultralight dark matter in neutrino oscillation experiments}",
    eprint = "2007.03590",
    archivePrefix = "arXiv",
    primaryClass = "hep-ph",
    reportNumber = "FERMILAB-PUB-20-260-T",
    doi = "10.1007/JHEP01(2021)094",
    journal = "JHEP",
    volume = "01",
    pages = "094",
    year = "2021"
}

@article{Losada:2021bxx,
    author = "Losada, Marta and Nir, Yosef and Perez, Gilad and Shpilman, Yogev",
    title = "{Probing scalar dark matter oscillations with neutrino oscillations}",
    eprint = "2107.10865",
    archivePrefix = "arXiv",
    primaryClass = "hep-ph",
    doi = "10.1007/JHEP04(2022)030",
    journal = "JHEP",
    volume = "04",
    pages = "030",
    year = "2022"
}

@article{Huang:2022wmz,
    author = "Huang, Guo-yuan and Lindner, Manfred and Mart{\'\i}nez-Mirav{\'e}, Pablo and Sen, Manibrata",
    title = "{Cosmology-friendly time-varying neutrino masses via the sterile neutrino portal}",
    eprint = "2205.08431",
    archivePrefix = "arXiv",
    primaryClass = "hep-ph",
    doi = "10.1103/PhysRevD.106.033004",
    journal = "Phys. Rev. D",
    volume = "106",
    number = "3",
    pages = "033004",
    year = "2022"
}

@article{Dev:2022bae,
    author = "Dev, Abhish and Krnjaic, Gordan and Machado, Pedro and Ramani, Harikrishnan",
    title = "{Constraining feeble neutrino interactions with ultralight dark matter}",
    eprint = "2205.06821",
    archivePrefix = "arXiv",
    primaryClass = "hep-ph",
    reportNumber = "FERMILAB-PUB-22-263-T",
    doi = "10.1103/PhysRevD.107.035006",
    journal = "Phys. Rev. D",
    volume = "107",
    number = "3",
    pages = "035006",
    year = "2023"
}

@article{Davoudiasl:2023uiq,
    author = "Davoudiasl, Hooman and Denton, Peter B.",
    title = "{Sterile neutrino shape shifting caused by dark matter}",
    eprint = "2301.09651",
    archivePrefix = "arXiv",
    primaryClass = "hep-ph",
    doi = "10.1103/PhysRevD.108.035013",
    journal = "Phys. Rev. D",
    volume = "108",
    number = "3",
    pages = "035013",
    year = "2023"
}

@article{Lopes:2023vxn,
    author = "Lopes, Il{\'\i}dio",
    title = "{Linking solar bosonic dark matter halos and active neutrinos}",
    eprint = "2310.14033",
    archivePrefix = "arXiv",
    primaryClass = "hep-ph",
    doi = "10.1103/PhysRevD.108.083028",
    journal = "Phys. Rev. D",
    volume = "108",
    number = "8",
    pages = "083028",
    year = "2023"
}

@article{Martinez-Mirave:2024dmw,
    author = "Mart{\'\i}nez-Mirav{\'e}, Pablo and Perez-Gonzalez, Yuber F. and Sen, Manibrata",
    title = "{Effects of neutrino-ultralight dark matter interaction on the cosmic neutrino background}",
    eprint = "2406.01682",
    archivePrefix = "arXiv",
    primaryClass = "hep-ph",
    reportNumber = "IPPP/24/28",
    doi = "10.1103/PhysRevD.110.055005",
    journal = "Phys. Rev. D",
    volume = "110",
    number = "5",
    pages = "055005",
    year = "2024"
}

@article{Goertz:2024gzw,
    author = "Goertz, Florian and Hager, Maya and Laverda, Giorgio and Rubio, Javier",
    title = "{Phasing out of darkness: from sterile neutrino dark matter to neutrino masses via time-dependent mixing}",
    eprint = "2407.04778",
    archivePrefix = "arXiv",
    primaryClass = "hep-ph",
    reportNumber = "IPARCOS-UCM-24-036",
    doi = "10.1007/JHEP02(2025)213",
    journal = "JHEP",
    volume = "02",
    pages = "213",
    year = "2025"
}

@article{Sahu:2025vyy,
    author = "Sahu, Purushottam and Sen, Manibrata",
    title = "{A Cosmic Amplification for Muon-to-Positron Conversion in Nuclei}",
    eprint = "2507.07176",
    archivePrefix = "arXiv",
    primaryClass = "hep-ph",
    journal = "",
    month = "7",
    year = "2025"
}

@article{DESI:2025zgx,
    author = "Abdul Karim, M. and others",
    collaboration = "DESI",
    title = "{DESI DR2 Results II: Measurements of Baryon Acoustic Oscillations and Cosmological Constraints}",
    eprint = "2503.14738",
    archivePrefix = "arXiv",
    primaryClass = "astro-ph.CO",
    reportNumber = "FERMILAB-PUB-25-0169-PPD",
    journal = "",
    month = "3",
    year = "2025"
}

@inproceedings{Smirnov:2004ju,
    author = "Smirnov, A. Yu.",
    title = "{Neutrinos: '...Annus mirabilis'}",
    booktitle = "{2nd International Workshop on Neutrino Oscillations in Venice (NO-VE 2003)}",
    eprint = "hep-ph/0402264",
    archivePrefix = "arXiv",
    pages = "1--21",
    month = "2",
    year = "2004"
}

@article{Minakata:2004xt,
    author = "Minakata, Hisakazu and Smirnov, Alexei Yu.",
    title = "{Neutrino mixing and quark-lepton complementarity}",
    eprint = "hep-ph/0405088",
    archivePrefix = "arXiv",
    doi = "10.1103/PhysRevD.70.073009",
    journal = "Phys. Rev. D",
    volume = "70",
    pages = "073009",
    year = "2004"
}

@article{Raidal:2004iw,
    author = "Raidal, Martti",
    title = "{Relation between the neutrino and quark mixing angles and grand unification}",
    eprint = "hep-ph/0404046",
    archivePrefix = "arXiv",
    doi = "10.1103/PhysRevLett.93.161801",
    journal = "Phys. Rev. Lett.",
    volume = "93",
    pages = "161801",
    year = "2004"
}

@inproceedings{Minakata:2005rf,
    author = "Minakata, Hisakazu",
    title = "{Quark-lepton complementarity: A Review}",
    booktitle = "{11th International Workshop on Neutrino Telescopes}",
    eprint = "hep-ph/0505262",
    archivePrefix = "arXiv",
    pages = "83--97",
    month = "5",
    year = "2005"
}

@article{Frampton:2004vw,
    author = "Frampton, P. H. and Mohapatra, R. N.",
    title = "{Possible gauge theoretic origin for quark-lepton complementarity}",
    eprint = "hep-ph/0407139",
    archivePrefix = "arXiv",
    reportNumber = "UMD-PP-05-002",
    doi = "10.1088/1126-6708/2005/01/025",
    journal = "JHEP",
    volume = "01",
    pages = "025",
    year = "2005"
}

@article{Ferrandis:2004vp,
    author = "Ferrandis, Javier and Pakvasa, Sandip",
    title = "{Quark-lepton complenmentarity relation and neutrino mass hierarchy}",
    eprint = "hep-ph/0412038",
    archivePrefix = "arXiv",
    reportNumber = "UH-511-1063-2004, LBNL-56686",
    doi = "10.1103/PhysRevD.71.033004",
    journal = "Phys. Rev. D",
    volume = "71",
    pages = "033004",
    year = "2005"
}

@article{Kang:2005as,
    author = "Kang, Sin Kyu and Kim, C. S. and Lee, Jake",
    title = "{Importance of threshold corrections in quark-lepton complementarity}",
    eprint = "hep-ph/0501029",
    archivePrefix = "arXiv",
    doi = "10.1016/j.physletb.2005.05.065",
    journal = "Phys. Lett. B",
    volume = "619",
    pages = "129--135",
    year = "2005"
}

@article{Antusch:2005ca,
    author = "Antusch, Stefan and King, Steve F. and Mohapatra, Rabindra N.",
    title = "{Quark-lepton complementarity in unified theories}",
    eprint = "hep-ph/0504007",
    archivePrefix = "arXiv",
    reportNumber = "SHEP-0510, UMD-PP-05-042",
    doi = "10.1016/j.physletb.2005.05.026",
    journal = "Phys. Lett. B",
    volume = "618",
    pages = "150--161",
    year = "2005"
}

@article{Schmidt:2006rb,
    author = "Schmidt, Michael A. and Smirnov, Alexei Yu.",
    title = "{Quark Lepton Complementarity and Renormalization Group Effects}",
    eprint = "hep-ph/0607232",
    archivePrefix = "arXiv",
    reportNumber = "TUM-HEP-639-06",
    doi = "10.1103/PhysRevD.74.113003",
    journal = "Phys. Rev. D",
    volume = "74",
    pages = "113003",
    year = "2006"
}

@article{Hochmuth:2006xn,
    author = "Hochmuth, Kathrin A. and Rodejohann, Werner",
    title = "{Low and High Energy Phenomenology of Quark-Lepton Complementarity Scenarios}",
    eprint = "hep-ph/0607103",
    archivePrefix = "arXiv",
    reportNumber = "MPP-2006-83, TUM-HEP-638-06",
    doi = "10.1103/PhysRevD.75.073001",
    journal = "Phys. Rev. D",
    volume = "75",
    pages = "073001",
    year = "2007"
}

@article{Plentinger:2007px,
    author = "Plentinger, Florian and Seidl, Gerhart and Winter, Walter",
    title = "{The Seesaw mechanism in quark-lepton complementarity}",
    eprint = "0707.2379",
    archivePrefix = "arXiv",
    primaryClass = "hep-ph",
    doi = "10.1103/PhysRevD.76.113003",
    journal = "Phys. Rev. D",
    volume = "76",
    pages = "113003",
    year = "2007"
}

@article{Altarelli:2009gn,
    author = "Altarelli, Guido and Feruglio, Ferruccio and Merlo, Luca",
    title = "{Revisiting Bimaximal Neutrino Mixing in a Model with S(4) Discrete Symmetry}",
    eprint = "0903.1940",
    archivePrefix = "arXiv",
    primaryClass = "hep-ph",
    reportNumber = "RM3-TH-09-4, CERN-PH-TH-2009-008, DFPD-09-TH-04",
    doi = "10.1088/1126-6708/2009/05/020",
    journal = "JHEP",
    volume = "05",
    pages = "020",
    year = "2009"
}

@article{Liu:1998nb,
    author = "Liu, Q. Y. and Mikheyev, S. P. and Smirnov, A. Yu.",
    title = "{Parametric resonance in oscillations of atmospheric neutrinos?}",
    eprint = "hep-ph/9803415",
    archivePrefix = "arXiv",
    reportNumber = "IC-98-30",
    doi = "10.1016/S0370-2693(98)01102-2",
    journal = "Phys. Lett. B",
    volume = "440",
    pages = "319--326",
    year = "1998"
}

@article{Akhmedov:1998ui,
    author = "Akhmedov, Evgeny K.",
    title = "{Parametric resonance of neutrino oscillations and passage of solar and atmospheric neutrinos through the earth}",
    eprint = "hep-ph/9805272",
    archivePrefix = "arXiv",
    reportNumber = "IC-98-43",
    doi = "10.1016/S0550-3213(98)00723-8",
    journal = "Nucl. Phys. B",
    volume = "538",
    pages = "25--51",
    year = "1999"
}

@article{Chizhov:1998ug,
    author = "Chizhov, M. and Maris, M. and Petcov, S. T.",
    title = "{On the oscillation length resonance in the transitions of solar and atmospheric neutrinos crossing the earth core}",
    eprint = "hep-ph/9810501",
    archivePrefix = "arXiv",
    reportNumber = "SISSA-53-98-EP",
    journal = "",
    month = "7",
    year = "1998"
}

@article{Akhmedov:1998xq,
    author = "Akhmedov, Evgeny K. and Dighe, A. and Lipari, P. and Smirnov, A. Y.",
    title = "{Atmospheric neutrinos at Super-Kamiokande and parametric resonance in neutrino oscillations}",
    eprint = "hep-ph/9808270",
    archivePrefix = "arXiv",
    reportNumber = "IC-98-98",
    doi = "10.1016/S0550-3213(98)00825-6",
    journal = "Nucl. Phys. B",
    volume = "542",
    pages = "3--30",
    year = "1999"
}

@article{Chizhov:1999he,
    author = "Chizhov, M. V. and Petcov, S. T.",
    title = "{Enhancing mechanisms of neutrino transitions in a medium of nonperiodic constant density layers and in the earth}",
    eprint = "hep-ph/9903424",
    archivePrefix = "arXiv",
    reportNumber = "SISSA-28-99-EP",
    doi = "10.1103/PhysRevD.63.073003",
    journal = "Phys. Rev. D",
    volume = "63",
    pages = "073003",
    year = "2001"
}

@article{Akhmedov:2006hb,
    author = "Akhmedov, Evgeny K. and Maltoni, Michele and Smirnov, Alexei Yu.",
    title = "{1-3 leptonic mixing and the neutrino oscillograms of the Earth}",
    eprint = "hep-ph/0612285",
    archivePrefix = "arXiv",
    doi = "10.1088/1126-6708/2007/05/077",
    journal = "JHEP",
    volume = "05",
    pages = "077",
    year = "2007"
}

@article{Kelly:2021jfs,
    author = "Kelly, Kevin J. and Machado, Pedro A. N. and Martinez-Soler, Ivan and Perez-Gonzalez, Yuber F.",
    title = "{DUNE atmospheric neutrinos: Earth tomography}",
    eprint = "2110.00003",
    archivePrefix = "arXiv",
    primaryClass = "hep-ph",
    reportNumber = "FERMILAB-PUB-21-459-T, NUHEP-TH/21-15",
    doi = "10.1007/JHEP05(2022)187",
    journal = "JHEP",
    volume = "05",
    pages = "187",
    year = "2022"
}

@article{Kelly:2019itm,
    author = "Kelly, Kevin James and Machado, Pedro AN and Martinez Soler, Ivan and Parke, Stephen J and Perez Gonzalez, Yuber F",
    title = "{Sub-GeV Atmospheric Neutrinos and CP-Violation in DUNE}",
    eprint = "1904.02751",
    archivePrefix = "arXiv",
    primaryClass = "hep-ph",
    reportNumber = "FERMILAB-PUB-19-136-T, NUHEP-TH/19-03",
    doi = "10.1103/PhysRevLett.123.081801",
    journal = "Phys. Rev. Lett.",
    volume = "123",
    number = "8",
    pages = "081801",
    year = "2019"
}

@article{T2K:2023smv,
    author = "Abe, K. and others",
    collaboration = "T2K",
    title = "{Measurements of neutrino oscillation parameters from the T2K experiment using $3.6\times 10^{21}$ protons on target}",
    eprint = "2303.03222",
    archivePrefix = "arXiv",
    primaryClass = "hep-ex",
    doi = "10.1140/epjc/s10052-023-11819-x",
    journal = "Eur. Phys. J. C",
    volume = "83",
    number = "9",
    pages = "782",
    year = "2023"
}

@article{NOvA:2021nfi,
    author = "Acero, M. A. and others",
    collaboration = "NOvA",
    title = "{Improved measurement of neutrino oscillation parameters by the NOvA experiment}",
    eprint = "2108.08219",
    archivePrefix = "arXiv",
    primaryClass = "hep-ex",
    reportNumber = "FERMILAB-PUB-21-373-ND",
    doi = "10.1103/PhysRevD.106.032004",
    journal = "Phys. Rev. D",
    volume = "106",
    number = "3",
    pages = "032004",
    year = "2022"
}

@article{DUNE:2020jqi,
    author = "Abi, B. and others",
    collaboration = "DUNE",
    title = "{Long-baseline neutrino oscillation physics potential of the DUNE experiment}",
    eprint = "2006.16043",
    archivePrefix = "arXiv",
    primaryClass = "hep-ex",
    reportNumber = "FERMILAB-PUB-20-251-E-LBNF-ND-PIP2-SCD, PUB-20-251-E-LBNF-ND-PIP2-SCD",
    doi = "10.1140/epjc/s10052-020-08456-z",
    journal = "Eur. Phys. J. C",
    volume = "80",
    number = "10",
    pages = "978",
    year = "2020"
}

@article{Kelly:2020fkv,
    author = "Kelly, Kevin J. and Machado, Pedro A. N. and Parke, Stephen J. and Perez-Gonzalez, Yuber F. and Funchal, Renata Zukanovich",
    title = "{Neutrino mass ordering in light of recent data}",
    eprint = "2007.08526",
    archivePrefix = "arXiv",
    primaryClass = "hep-ph",
    reportNumber = "FERMILAB-PUB-20-330-T",
    doi = "10.1103/PhysRevD.103.013004",
    journal = "Phys. Rev. D",
    volume = "103",
    number = "1",
    pages = "013004",
    year = "2021"
}

@article{DayaBay:2022orm,
    author = "An, F. P. and others",
    collaboration = "Daya Bay",
    title = "{Precision Measurement of Reactor Antineutrino Oscillation at Kilometer-Scale Baselines by Daya Bay}",
    eprint = "2211.14988",
    archivePrefix = "arXiv",
    primaryClass = "hep-ex",
    doi = "10.1103/PhysRevLett.130.161802",
    journal = "Phys. Rev. Lett.",
    volume = "130",
    number = "16",
    pages = "161802",
    year = "2023"
}

@article{Nunokawa:2005nx,
    author = "Nunokawa, Hiroshi and Parke, Stephen J. and Zukanovich Funchal, Renata",
    title = "{Another possible way to determine the neutrino mass hierarchy}",
    eprint = "hep-ph/0503283",
    archivePrefix = "arXiv",
    reportNumber = "FERMILAB-PUB-05-041-T",
    doi = "10.1103/PhysRevD.72.013009",
    journal = "Phys. Rev. D",
    volume = "72",
    pages = "013009",
    year = "2005"
}

\end{document}

\sloppy  

\preprint{IFT-UAM/CSIC-26-02, TIFR/TH/26-5}

\title{Supplementary Material: Diagonalizing the net Hamiltonian}

\author{Susobhan Chattopadhyay}%\,\orcidlink{0000-0001-7948-4332}}
\email{susobhan.chattopadhyay@tifr.res.in}
\affiliation{Tata Institute of Fundamental Research,
Homi Bhabha Road, Mumbai, 400005, India}

\author{Yuber F. Perez-Gonzalez}%\,\orcidlink{0000-0002-2020-7223}}
\email{yuber.perez@uam.es}
\affiliation{Departamento de F\'{i}sica Te\'{o}rica and Instituto de F\'{i}sica Te\'{o}rica (IFT) UAM/CSIC, Universidad Aut\'{o}noma de Madrid, Cantoblanco, 28049 Madrid, Spain}

\author{Manibrata Sen}%\,\orcidlink{0000-0001-7948-4332}}
\email{manibrata@iitb.ac.in}
\affiliation{Department of Physics, Indian Institute of Technology Bombay, Powai, Maharashtra 400076 India}

\begin{abstract}
    
\end{abstract}

\maketitle

In the scenario considered in the paper, neutrinos can interact with ULDM $\phi$ through mediators $\chi$ via the Lagrangian
\begin{eqnarray}
 \mathcal{L} \subset \sum_{\substack{\alpha=e,\mu,\tau \\ k=1,2}}\biggl([M_{\rm vac}]_{\alpha\beta}\,\overline{(\nu_{\alpha,L})^C}\,\nu_{\beta,L}  + g_{\alpha k}\,\overline{(\chi_{k,L})^C}\,\nu_{\alpha,L}\,\phi^\star 
   +  m_{\chi_k}\,\overline{(\chi_{k,L})^C}\,\chi_{k,L} + \text{h.c.} + V(\phi)\biggr)\,,
\end{eqnarray}
while $g$ denotes the interaction couplings, $m_\chi$ are the masses of the mediators and $V(\phi)$ is the self potential of the ULDM. The major difference with~\cite{Sen:2023uga} is that neutrinos are assumed to have tiny masses, governed by the mass matrix, $M_{\rm vac} =[V_{\rm CKM}]\cdot\,{\rm diag} (m_1,m_2,m_3)\cdot\,[V_{\rm CKM}^\dagger]\,$, appropriately rotated by the CKM mixing matrix.

Let us start with the simpler scenario as in \cite{Sen:2023uga,Chattopadhyay:2025ccy} with vanishing vacuum masses, i.e. $[M_{\text{vac}}]_{3\times3} = 0$. In this scenario, the $5\times5$ mass matrix in the basis $\{\nu_e,\nu_\mu, \nu_\tau, \chi_1, \chi_2\}$ may be written as 
\begin{equation}\label{eq:mass_mat}
    \mathbb{M} = \begin{pmatrix}
        0_{3\times3} & \ \ & [g]_{\alpha k}\,\langle\phi\rangle^\star_{\text{coh}}\\
        [g^T]_{k \alpha}\,\langle\phi\rangle^\star_{\text{coh}} & \ \ & [m_{\chi_k}]_D
    \end{pmatrix}\,,
\end{equation}
where $\langle \phi \rangle$ is the expectation value of the coherent ULDM scalar field, parameterized as $\langle \phi \rangle^*_{\rm coh}=F e^{i \Phi}$~\cite{Sen:2023uga}, where $F$ is the amplitude and $\Phi$ is a phase dependent on the ULDM mass. Also, $[m_{\chi_k}]_D$ refers to $2\times2$ diagonal matrix characterising the masses of the mediators, $m_{\chi_{1,2}}$. 
We thus obtain the effective Hamiltonian for propagation to be
\begin{equation}
    \mathbb{H} = \frac{1}{2E}\mathbb{M}\mathbb{M^\dagger} = \frac{1}{2E} \begin{pmatrix}
        F^2\, [g] \cdot [g^\dagger] & \ \ \ \  & F\, e^{-i\Phi} [g]\cdot [m_{\chi_k}]_D \\
        F\, e^{i\Phi} [m_{\chi_k}]_D \cdot [g^\dagger]  & \ \ \ \ & F^2\,[g^T] \cdot [g^\star] + [m_{\chi_k}^2]_D
        \end{pmatrix}\,. \label{eq:hamltonian_H}
\end{equation}
This results in the following evolution equations, 
\begin{eqnarray}
    2E\left(i\frac{d}{d t}\lvert\nu_\alpha\rangle\right) &=& \sum_{\beta=e,\mu,\tau}\sum_{j=1,2} F^2 g_{\alpha j}g_{\beta j}^\star\lvert\nu_\beta\rangle + \sum_{j=1,2} F g_{\alpha j} m_{\chi_j}e^{-i\Phi}\lvert\chi_j\rangle\,,\\
    2E\left(i\frac{d}{d t}\lvert\chi_k\rangle\right) &=& \sum_{\beta=e,\mu,\tau} Fg_{\beta k}^\star m_{\chi_k}e^{i\Phi}\lvert\nu_\beta\rangle + \sum_{j=1,2}\left(\sum_{\beta=e,\mu,\tau}F^2g_{\beta k}^\star g_{\beta j} + m_{\chi_j}^2\delta_k^j\right)\lvert\chi_j\rangle\,.
\end{eqnarray}
We follow the similar procedure as in~\cite{Chattopadhyay:2025ccy} to get rid of the time-dependent complex phase $\Phi$ from the Hamiltonian by redefining the $\lvert \chi_k\rangle \to e^{-i\Phi} \lvert\chi_k\rangle$ to get rid of the complex phase $\Phi$ and work in this new ``tilde" basis. The states in the tilde basis are
\begin{equation}
\lvert \widetilde{\nu}_\alpha \rangle = \left\{\lvert\nu_e\rangle,\lvert\nu_\mu\rangle,\lvert\nu_\tau\rangle,\lvert\widetilde{\chi}_1\rangle,\lvert\widetilde{\chi}_2\rangle\right\} = \left\{\lvert\nu_e\rangle,\lvert\nu_\mu\rangle,\lvert\nu_\tau\rangle,e^{-i\Phi}\lvert\chi_1\rangle,e^{-i\Phi}\lvert\chi_2\rangle\right\}.\label{eq:tilde_basis}
\end{equation}
Eventually, the Hamiltonian in this basis becomes
\begin{equation}
    \widetilde{\mathbb{H}} = \frac{1}{2E} \begin{pmatrix}
        F^2 [g] \cdot [g^\dagger] & \ \ \ \ \ \ \ \  & F [g]\cdot [m_{\chi_k}]_D \\
        F [m_{\chi_k}]_D \cdot [g^\dagger] & \ \ \ \ \ \ \ \  & F^2[g^T] \cdot [g^\star] + [m_{\chi_k}^2+ 2E\dot{\Phi}]_D
        \end{pmatrix}\label{eq:tilde_basis_H}\,.
\end{equation}
We diagonalize this Hamiltonian in a two step procedure. First, we make a Singular Value Decomposition (SVD) of the matrix $[g]_{3\times2}$ as
\begin{equation}
    [U^\dagger]_{3\times3} \cdot F [g]_{3\times2} \cdot [W]_{2\times2} = \begin{pmatrix}
        0 & 0\\
        m_{a_1} & 0\\
        0 & m_{a_2}
    \end{pmatrix}\,,\label{eq:SVD}
\end{equation}
where $[U]_{3\times3}$ is a $3\times3$ unitary matrix and $[W]_{2\times2}$ is an orthogonal $2\times 2$ matrix. It can be shown \cite{Chattopadhyay:2025ccy} that the experimentally observable oscillation probabilities are independent of the matrix $[W]_{2\times2}$.
For simplicity of the analysis here, let us assume $m_{\chi_1} = m_{\chi_2} \equiv m_\chi$. Substituting eq. \eqref{eq:SVD} in eq. \eqref{eq:tilde_basis_H}, we get
\begin{eqnarray}
     \mathbb{H'} &=& \begin{pmatrix}
        [I]_{3\times3} & 0\\
        0 & [W^T]_{2\times2}
    \end{pmatrix} \cdot \begin{pmatrix}
        [U^\dagger]_{3\times3} & 0\\
        0 & [I]_{2\times2}
    \end{pmatrix} \cdot \widetilde{\mathbb{H}} \cdot \begin{pmatrix}
        [U]_{3\times3} & 0\\
        0 & [I]_{2\times2}
    \end{pmatrix} \cdot \begin{pmatrix}
        [I]_{3\times3} & 0\\
        0 & [W]_{2\times2}
    \end{pmatrix}\nonumber\\
    &=& \frac{1}{2E}\begin{pmatrix}
    0 & 0 & 0 & 0 & 0\\
    0 & m_{a_1}^2 & 0 & m_{a_1} m_\chi & 0\\
    0 & 0 & m_{a_2}^2 & 0 & m_{a_2} m_\chi\\
    0 & m_{a_1} m_\chi & 0 & m_{a_1}^2+ m_{\chi}^2 + 2E\dot{\Phi} & 0\\
    0 & 0 & m_{a_2} m_\chi  & 0 & m_{a_2}^2+ m_{\chi}^2 + 2E\dot{\Phi}
    \end{pmatrix}\,.
\end{eqnarray}

Thus, one of the neutrino species is completely decoupled and the rest are decoupled as pairs of two neutrinos with the decoupled sub-blocks being
\begin{eqnarray}
    \begin{pmatrix}
    H'_{22} & H'_{24}\\
    H'_{42} & H'_{44}
\end{pmatrix} = \begin{pmatrix}
    m_{a_1}^2 & m_{a_1} m_\chi\\
    m_{a_1} m_\chi & m_{a_1}^2+ m_{\chi}^2 + 2E\dot{\Phi}
\end{pmatrix}~,\\
\begin{pmatrix}
    H'_{33} & H'_{35}\\
    H'_{53} & H'_{55}
\end{pmatrix} = \begin{pmatrix}
    m_{a_2}^2 & m_{a_2} m_\chi\\
    m_{a_2} m_\chi & m_{a_2}^2+ m_{\chi}^2 + 2E\dot{\Phi}
\end{pmatrix}~,
\end{eqnarray}
As a next step, we diagonalize these two sub-blocks with the rotation matrices $R_{22'}(\alpha_{22'})$ and $R_{33'}(\alpha_{33'})$, respectively, with the mixing angles given by
 \begin{equation}
     \alpha_{22'} = \frac{1}{2} \text{tan}^{-1}\left(\frac{2m_{a_1}m_\chi}{m_\chi^2 + 2 E \dot{\Phi}}\right)\ \  \text{and}\ \ \alpha_{33'} = \frac{1}{2} \text{tan}^{-1}\left(\frac{2m_{a_2}m_\chi}{m_\chi^2 + 2 E \dot{\Phi}}\right)~.
 \end{equation}
The final mass eigenstates are labeled as $\{\nu_1,\nu_2,\nu_3,\nu_{2'},\nu_{3'}\}$. 
 
Eventually, we have diagonalized $\widetilde{\mathbb{H}}$ as
\begin{equation}
    \mathbb{H}_D =  \mathbb{P}^\dagger \cdot \widetilde{\mathbb{H}}  \cdot \mathbb{P}~,
\end{equation}
where $\mathbb{H}_D$ is a diagonal matrix and
\begin{equation}
    \mathbb{P} =   \begin{pmatrix}
        [U]_{3\times3} & 0\\
        0 & [I]_{2\times2}
    \end{pmatrix} \cdot \begin{pmatrix}
        [I]_{3\times3} & 0\\
        0 & [W]_{2\times2}
    \end{pmatrix} \cdot R_{22'}(\alpha_{22'})\cdot R_{33'}(\alpha_{33'})~.\label{eq:P_defn}
\end{equation} 
The final diagonalized mass-squared matrix, $\mathbb{M}_D^2 = 2 E\cdot\mathbb{H}_D$, becomes
    \begin{eqnarray}
        \mathbb{M}_D^2 = \text{diag}\left\{0,\,\overline{m^2}_{\!22'}-\frac{\Delta m_{22'}^2}{2},\,\overline{m^2}_{\!33'}-\frac{\Delta m_{33'}^2}{2},\,\overline{m^2}_{\!22'}+\frac{\Delta m_{22'}^2}{2},\,\overline{m^2}_{\!33'}+\frac{\Delta m_{22'}^2}{2}\right\}\,,
    \end{eqnarray}
where 
\begin{equation}
    \overline{m^2}_{\!22',33'} = m_{a_{1,2}}^2+\frac{m_\chi^2+2E\dot{\Phi}}{2}~,    
\end{equation}
and 
\begin{equation}
\Delta m_{22',33'}^2 = \sqrt{4 m_{a_{1,2}}^2 m_\chi^2+\left(m_\chi^2+2E\dot{\Phi}\right)^2}~.    
\end{equation}
 
 The measured values of $\Delta m^2_\text{sol}$ and $\Delta m^2_\text{atm}$ may now be identified with
 \begin{equation}
     \Delta m^2_\text{sol} \approx \Delta m^2_{12} = \left( \overline{m^2}_{\!22'}-\frac{\Delta m_{22'}^2}{2} \right)~,\ \ \ \ \Delta m^2_\text{atm} \approx \Delta m^2_{31} = \left( \overline{m^2}_{\!33'}-\frac{\Delta m_{33'}^2}{2}\right)~.
 \end{equation}
 When $\alpha_{22',33'} \ll 1$, the matrix $[U]_{3\times3}$ diagonalizes the active $3\times3$ sub-matrix of $\widetilde{\mathbb{H}}$. Hence, we may identify $U_{\text{PMNS}} = [U]_{3\times3}$. Note that this identification is also valid when $\Delta m_{22',33'}^2 \ll \overline{m^2}_{\!22',33'}$.

Focusing on the scenario described in this paper, we perturb the $[0]_{3\times3}$ block in $\mathbb{M}$ by a small $M_{\rm vac}$ such that 
\begin{equation}
M_{\rm vac} =[V_{\rm CKM}]\cdot\,{\rm diag} (m_1,m_2,m_3)\cdot\,[V_{\rm CKM}^\dagger]\,.
\end{equation}
The Hamiltonian in the tilde basis is modified as
\begin{align}
    \widetilde{\mathbb{H}} =& \frac{1}{2E} \begin{pmatrix}
        U_{\rm PMNS} \cdot \Delta M_{\rm D}^2 \cdot U_{\rm PMNS}^\dagger &\qquad  U_{\rm PMNS} \cdot \Delta M_{\rm D} \cdot \tilde{m}_{\chi_k} \\
        \tilde{m}_{\chi_k}^T \cdot \Delta M_{\rm D}^T \cdot U_{\rm PMNS}^\dagger &\qquad [\Delta M_D^2]_{2\times 2} + [m_{\chi_k}^2+ 2E\dot{\Phi}]_D
        \end{pmatrix} \notag\\
        &+ \frac{1}{2E} \begin{pmatrix}
            M_{\rm vac} \cdot M_{\rm vac}^{\dagger} &\ & e^{2i\Phi}\, M_{\rm vac} \cdot U_{\rm PMNS}^* \cdot \Delta M_{\rm D} \cdot W^\dagger \\
            e^{-2i\Phi}\, W\cdot \Delta M_{\rm D}^T \cdot U_{\rm PMNS}^T  \cdot M_{\rm vac}^{\dagger} &\ & [0]_{2\times2}
        \end{pmatrix}\,,
        \label{eq:modifiedHamil}
\end{align}
where 
\begin{equation}\label{eq:DMD}
     \Delta M_{\rm D} \equiv \begin{pmatrix}
        0 & 0\\
        \sqrt{\Delta m^2_\text{sol}} & 0\\
        0 & \sqrt{\Delta m^2_\text{atm}}
    \end{pmatrix}\,,\quad \quad \Delta M_{\rm D}^2 \equiv  \Delta M_{\rm D}\cdot \Delta M_{\rm D}^T = {\rm diag}(0, \Delta m^2_\text{sol}, \Delta m^2_\text{atm})\, ,
\end{equation}
and $[\Delta M_D^2]_{2\times 2} = \Delta M_{\rm D}^T\cdot \Delta M_{\rm D} =  {\rm diag}(\Delta m^2_\text{sol}, \Delta m^2_\text{atm})$ refers to the bottom-right $2\times2$ block of $\Delta M_{\rm D}^2$.
Clearly when $F = 0$, $V_{\rm CKM}$ diagonalizes the $3\times3$ active sector but for non-zero $F$, we have $U \approx U_{\rm PMNS}$ when $M_{\rm vac} \ll F[g]$. Since the formalism is similar to the ones discussed in~\cite{Sen:2023uga,Chattopadhyay:2025ccy}, the bounds on the couplings required for this mapping are also unchanged.

There is an additional subtlety when we proceed to calculate oscillation probabilities using the modified Hamiltonian eq. \eqref{eq:modifiedHamil}. The second term (perturbation) in eq. \eqref{eq:modifiedHamil} is time-dependent as $\Phi \sim m_\phi t$. Further, eq. \eqref{eq:modifiedHamil} is periodic for $\Phi \to \Phi+\pi$. Thus, we evaluate the oscillation probability by discretizing the time into small steps and then evolving the initial neutrino state as
\begin{equation}
    P_{\alpha \beta} (\Phi_0,E) =  \left\lvert \left\langle \nu_\beta\left\lvert \prod_{n=0}^{N-1} \left[ \mathcal{U}_M\left(\Phi_0,n\Delta x\right) \cdot Diag\left(\exp\left(-\frac{\Delta x \, m_M^2\left(\Phi_0,n\Delta x\right)}{2E}\right)\right)\cdot \mathcal{U}_M^\dagger\left(\Phi_0,n\Delta x\right)\right]\right\rvert\nu_\alpha \right\rangle \right\rvert^2
\end{equation}
where $\Delta x = L/N$. After this, we take the average over the initial phase $\Phi_0$. This is because during the large timescale neutrino experiments, the neutrinos will scan over all possible initial phases. Therefore,
\begin{equation}
    \langle P_{\alpha\beta} (E) \rangle = \frac{1}{\pi}\int_{0}^\pi d\Phi_0\, P_{\alpha\beta}(\Phi_0,E) 
\end{equation}

%%%%%%%%%%%%%%%%%%%%%%%%%%%%%%%%%%%%%%%%%%%%%%%%%%%%%%%%%%%%%%%%%%%%%%%%%%%%%%
\section{Redshift dependence of neutrino masses and mixing}
%%%%%%%%%%%%%%%%%%%%%%%%%%%%%%%%%%%%%%%%%%%%%%%%%%%%%%%%%%%%%%%%%%%%%%%%%%%%%%

In our scenario, masses and mixing depend on the epoch and position at which neutrinos are emitted. Let us consider a simplified two flavour scenario to understand the main effects. Considering the active sub-sector in Eq.~\eqref{eq:modifiedHamil}, we have
\begin{align}\label{eq:modifiedHamil_2f}
    \hat{\mathbb{H}} = \frac{1}{2E} U (\theta_\chi) \cdot \begin{pmatrix}
        0 & 0\\
        0 & \Delta m^2 (z)
    \end{pmatrix}\ \cdot U (\theta_\chi)^\dagger + \frac{1}{2E} U (\theta_0) \cdot \begin{pmatrix}
        m_1^2 & 0\\
        0 & m_2^2
    \end{pmatrix} \cdot U (\theta_0)^\dagger\,,
\end{align}
where, as usual,
\begin{align}
    U (\theta) = \begin{pmatrix}
        \cos\theta & \sin\theta\\
        -\sin\theta & \cos\theta
    \end{pmatrix}\,,
\end{align}
$\theta_\chi, \theta_0$ refer to mixing angles in the DM background and vacuum, respectively and $\Delta m^2(z)$ is the quadratic mass splitting from refractive effects depending on redshift. From Ref.~\cite{Sen:2023uga}, for the cosmic neutrino background (C$\nu$B), we have that
\begin{align}
    \Delta m^2(z) = \xi\,\Delta m^2(0)\, (1+z)^5\,\frac{E_0^2}{E_0^2(1+z)^2 - E_R^2},
\end{align}
with $\Delta m^2(0)$ the quadratic mass difference today, $\xi \approx 10^{-5}$, value related to the local DM overdensity near Earth, $E_0$ the C$\nu$B energy today, and $E_R$ is the resonance energy for the neutrino-DM interaction.
From this, we observe that the neutrino mass for the C$\nu$B grows in the Early Universe, while today the C$\nu$B might be dominated by the vacuum masses $m_{12}^2$ and mixings. Choosing for simplicity $m_1=0$ and denoting $\delta m^2\equiv m_2^2$, the mass matrix squared is\footnote{We do not consider the Hamiltonian from Eq.~\eqref{eq:modifiedHamil_2f} directly because for the C$\nu$B one would need to analyse the whole Dirac Hamiltonian instead of its ultra-relativistic limit.}
\begin{align}
    \hat{\mathbb{M}}^2= 
    \begin{pmatrix}
        \Delta m^2 (z) \sin^2\theta_\chi + \delta m^2 \sin^2\theta_0 & \Delta m^2 (z) \sin\theta_\chi\cos\theta_\chi + \delta m^2 \sin\theta_0\cos\theta_0\\
        \Delta m^2 (z) \sin\theta_\chi\cos\theta_\chi + \delta m^2 \sin\theta_0\cos\theta_0& \Delta m^2 (z) \cos^2\theta_\chi + \delta m^2 \cos^2\theta_0 
    \end{pmatrix} \,,
\end{align}
Diagonalizing one finds the full mass eigenstates squared and the mixing angles,
\begin{subequations}
    \begin{align}
        Dm_{1,2}^2(z) &= \frac{1}{2}\left(\delta m^2+\Delta m^2 (z) \pm \sqrt{(\delta m^2)^2+(\Delta m^2 (z))^2 + 2 \delta m^2\Delta m^2 (z)\cos(2(\theta_\chi - \theta_0))}\right),\\
        \sin^2 2\theta_z &= \frac{(\Delta m^2 (z) \sin 2\theta_\chi + \delta m^2 \sin 2\theta_0)^2}{(\delta m^2)^2+(\Delta m^2 (z))^2 + 2 \delta m^2\Delta m^2 (z)\cos(2(\theta_\chi - \theta_0))}.
    \end{align}
\end{subequations}
Interestingly, for
\begin{align}
    \Delta m^2 (z) \cos 2\theta_\chi =- \delta m^2 \cos 2\theta_0,
\end{align}
$\sin^2 2\theta_z = 1$, so that the effective mixing at $z$ becomes maximal, i.e., we would have a resonance.
Since we have that $\delta m^2>0$, we require $(\theta_\chi > \pi/4 \land \theta_0 < \pi/4) \lor (\theta_\chi < \pi/4 \land \theta_0 > \pi/4)$ for $\Delta m^2 (z)>0$, with opposite conditions for $\Delta m^2 (z)<0$.

\begin{figure}[!t]
    \centering
    \includegraphics[width=0.9\linewidth]{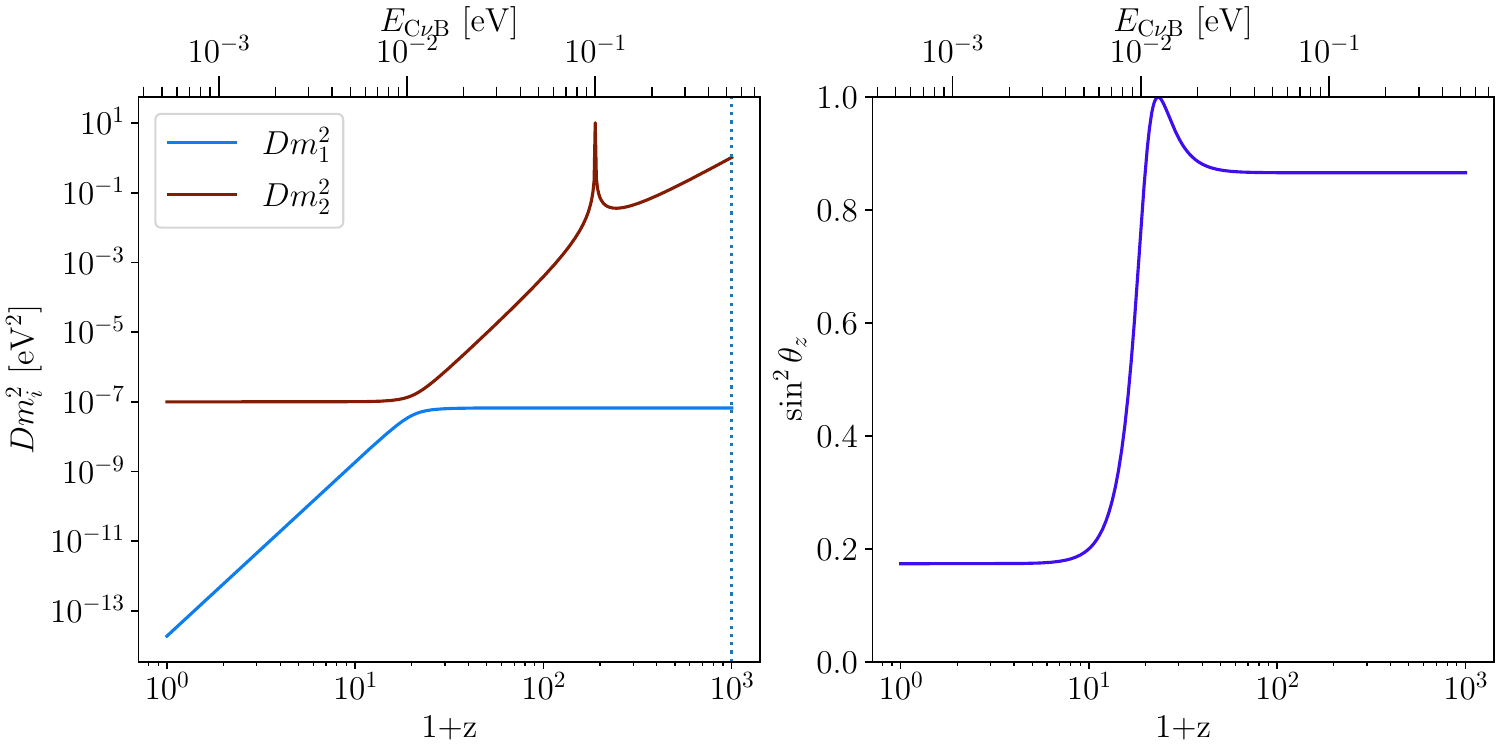}
    \caption{Mass-squared eigenvalues (left) and effective mixing angle, parametrized via $\sin^2\theta_z$ (right) as function of redshift for the cosmic neutrino background. We consider the values $\delta m^2 = 10^{-7}~{\rm eV^2}, \Delta m^2 (0) = 10^{-4}~{\rm eV^2}, \theta_0 = 5^\circ, \theta_\chi = 60^\circ, E_0 = 0.54~{\rm meV}$, and $E_R = 0.1~{\rm eV}$ for illustration purposes.} \label{fig:redshift_2f}
\end{figure}

As illustration, we present the redshift evolution of the mass-squared eigenvalues $Dm_{12}^2(z)$ (left) and the effective mixing angle (right) assuming $\delta m^2 = 10^{-7}~{\rm eV^2}, \Delta m^2 (0) = 10^{-4}~{\rm eV^2}, \theta_0 = 5^\circ, \theta_\chi = 60^\circ, E_0 = 0.54~{\rm meV}, E_R = 0.1~{\rm eV}$ in Fig.~\ref{fig:redshift_2f}.
In this example we observe the resonance effect around $z\sim 20$, point at which the mixing becomes maximal and there is a level crossing in the mass eigenstates. Additionally, for the eigenvalue $D m_1^2$ we have the resonance effect at $z\sim 200$ coming from the $\chi$-exchange in the s-channel of the neutrino-DM interaction at $E_\nu = E_R$, see Ref.~\cite{Sen:2023uga}.
%============================================%
%============================================%

\bibliography{biblio}